\documentclass[prb,aps,showpacs,unsortedaddress,superscriptaddress,amsmath,amssymb,twocolumn,floatfix]{revtex4}
\usepackage{epsfig}
\usepackage{epstopdf}
\usepackage{xfrac}
\usepackage{graphicx}
\usepackage{dcolumn}
\usepackage{bm}
\usepackage{subcaption}
\usepackage{booktabs}
\usepackage{blindtext}
\begin{document}
\title{Role of spin-orbit coupling effects in rare-earth metallic tetra-borides : a first principle study}
\author{Ismail Sk}
\affiliation{Bajkul Milani Mahavidyalaya, Purba Medinipur-721655,West Bengal, India}
\affiliation{Kazi Nazrul University, Asansol-713340,West Bengal, India}
\author{Nandan Pakhira}
\affiliation{Kazi Nazrul University, Asansol-713340,West Bengal, India}
\begin{abstract}
	We have investigated the electronic structure of rare-earth tetraborides, $\textrm{RB}_{4}$, using first-principle electronic structure methods (DFT) implemented in Quantum Espresso (QE). 
In this article we have studied heather-to neglected strong spin-orbit coupling (SOC) effects present in these systems on the electronic structure of these system in the non-magnetic ground state. 
The calculations were done under GGA and GGA+SO approximations using ultrasoft pseudopotentials and fully relativistic ultrasoft pseudopotentials (for SOC case). Perdew-Burke-Ernzerhof generalized 
gradient approximation (PBE-GGA) exchange-correlation functionals within the linearized plane-wave (LAPW) method as implemented in QE were used. The projected density of states consists of 3 distinct 
spectral peaks well below the Fermi energy and separated from the continuum density of states around the Fermi energy. The discrete peaks arises due to rare-earth $s$-orbital, rare-earth $p$ + B $p$ 
and B $p$-orbitals while the continuum arises due to hybridized B $p$, rare-earth $d$ orbitals. Upon inclusion of SOC the peak arising due to rare-earth $p$-orbitals gets split into two peaks 
corresponding to $j=0.5$ and $j=1.5$ configurations. In case of $\textrm{LaB}_{4}$, in the presence of SOC, spin-split $4f$ orbitals contributes to density of states at the Fermi level while the 
density of states at the Fermi level largely remains unaffected for all other materials under consideration. 
\end{abstract}
\maketitle
\section{Introduction}
The strong Coulomb correlations present in $3d$ and $4d$ transition metal compounds as well as in $4f$ lanthanides and $5f$ actinides are key to understanding nonvel and exotic properties. 
The rare earth lanthanides except Pm are good conductors of heat and electricity. Pm is radioactive with very short life and its occurance in nature is extremely rare. The rare-earth metallic 
tetra-borides exhibit various valency such as di, tri and tetravalent state [1]. Cerium (Ce) and Terbium (Tb) primarily show tetravalent state where as the other metallic tetraborides mostly 
show trivalent state [1]. Recently intermediated valance state of Yb between $\textrm{Yb}^{2+}$ and $\textrm{Y}b^{3+}$ is experimentally observerd and Kondo interaction is significant in this system [2].

Recent observation of fractional magnetic plateau in $\textrm{TmB}_{4}$ and $\textrm{ErB}_{4}$ have created lot of interest in these class of materials. Stable magnetization plateau occuring at 
1/2 fraction (of saturation magnetization) and fractional plateaus at $1/7, 1/8, \cdots$ etc [3,4]. fractions are similar to the plateaus observed in the Hall resistivity of two dimensional degenerate 
electron gases subject to a perpendicular magnetic field.

It is interesting to mention that the position of the rare-earth atoms as shown in Fig. 1 forms a two dimensional Archimedian Shastry-Sutherland lattice (SSL) [5]. SSL consisting of localized 
spin-1/2 is an example of geometrically frustrated system with huge spin degeneracy and the observation of magnetization plateaus is often attributed to this degenearacy. Insulating 
$\textrm{SrCu}_{2}(\textrm{BO}_{3})$ [6] is a well studied system which can be effectively mapped onto a nearest-neighbour SSL. However in metallic rare-earth tetraborides ``localized'' spins 
interacts only through long range RKKY [7] type of interactions. Hence the mapping of interacting fermionic model onto an effective spin-1/2 models on SSL with nearest neighbour interaction is 
highly non-trivial [5,8]. Correlated and frustrated systems are of great academic interest as well as they have many potential technologicl applications like memory device, spintronics, quantum 
computation etc. [9]. The very first step towards understanding the intriguing thermodynamic and transport properties in these complex systems is to study their electronic band structure. 
In an earlier work [10] electronic structure of $\textrm{RB}_{4}$ (except $\textrm{TmB}_{4}$) have been studied using first principle method. However strong atomic spin-orbit coupling 
effects present in rare-earth atoms have been neglected. Inclusion of SOC for certain systems [R= Yb, Pr, Gd, Tb, Dy] in the mangetic state have been considered and also considered of $4f$ for 
$\textrm{RB}_{4}$ but there is no systematic study of such effects in the non-magnetic (paramagnetic) state [11]. In the present work we make a detailed study of SOC effects on the electronic structure of 
rare earth tetra-borides. In particular we have chosen systems (R=La, Ce, Nd, Sm) with relatively low SOC effects as well as systems (R=Ho,Er,Tm,Lu) with relatively high SOC effects.

The organization of the rest of the paper is as follows. In Sec. II we discuss crystal structure of the system. In Sec. III we elaborate the computational details for band structure. In Sec. IV we 
discuss the results for various systems and finally in Sec. V we conclude.
\section{Crystal structure}
$\textrm{RB}_{4}$ crystallizes in the tetragonal symmetrey with space group P4/mbm [12,13]. Fig. 1 summarizes crystal structure of $\textrm{RB}_{4}$ from different perspectives. Fig. 1 (a) displays the 
full tetragonal structure which consists of alternate layers of rare-earth (R) and B ions stacked along $c$-axis. Fig. 1(b) shows the top view of the crystal structure. There are two distinct 
types of B atoms - (i) planar and (ii) octahedral. Boron atoms form octahedra as well as 7-atom rings in the $a-b$ plane [14]. Ring forming planar B atoms (shown in blue) which are not part of 
octahedra also forms dimers and these dimers are aranged in a regular pattern. In Fig. 1(c) we show one unit cell formed by four such B octahedra. In Fig. 1(d) we show SSL formed by the B atoms. 
From Fig. 1(b) it is clear that out of the 4 B atoms two are nearer than other two. The exchange interaction between the two near B atoms mimics the nearest neighbour interaction (J) and the 
interaction between the distant B atoms mimics interaction along alternate diagonals. It is important to mention that B atoms play a crucial role in the electronic structure of these systems as 
they are in the $sp$-hybridized state.

\begin{figure}[htbp]
\includegraphics[scale=0.4]{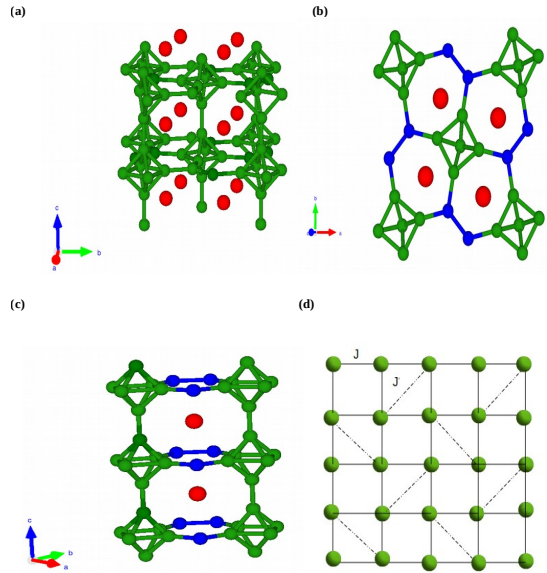} 
	\caption{(Color online) Tetragonal crystal structure of $RB_{4}$. Fig. 1(a) represents the full structure consisting of different layers of rare-earth, R, ions (red) and B (green) stacked 
	along $c$-axis. (b) Top view of the B sublattice (in the $a-b$ plane) comprising of 7 atom ring and a square formed by the position of the R atoms. (c) Side view of the B sublattice 
	(along $c$-axis) showing two different types of B ; one forming dimer (shown in blue) and the other part of the B octahedra (shown in green). (d) Shastry-Sutherland lattice in two dimension.}
\end{figure}
\section{Computational Details}
First-principle calculations were performed using density functional theory (DFT)[15,16] as implemented in the open source package Quantum Espresso [17] under the Burai [18] framework. 
The calculations are done within GGA and GGA+SO approximation. We have used Ultra soft pseudo-potentials [19] Marzari-
Vanderbit smaering [20] for structural optimization and total energy calculation of the system. 
Further, Perdew-Burke-Ernzerhof Generalized Gradient Approximation (PBE-GGA) exchange-correlation functional within the linearized augmented plane wave (LAPW) method is employed [21,22]. 
For the case with SOC effect full relativistic Ultra soft pseudo-potentials were used. The total Hamiltonian of the Kohn-sham DFT calculations with Spin-Orbit coupling can be 
written as [23]
\begin{equation}
\hat{H} = \hat{T} + \hat{V}_{ext} + \hat{V}_{es} + \hat{V}_{xc} + \hat{H}^{SOC}
           = \hat{T} + \hat{V}_{a} + \hat{H}^{SOC},
\end{equation}
where, $\hat{T}$,$\hat{V}_{ext}$,$\hat{V}_{es}$,$\hat{V}_{xc}$ and $\hat{H}^{SOC}$ are the kinetic energy operator, external potential operator, electrostatic or hartree potential operator, exchange-correlation potential operator and spin-orbit coupling operator respectively. $\hat{V}_{a}$ is the applied field or Kohn-Sham potential operator.
The Hamiltonian $\hat{H}^{SOC}$ for relativistic limit in terms of momentum and spin operator can be expressed as [23]
\begin{equation}
\hat{H}^{SOC}= \frac{i}{4c^2}({\nabla}\hat{V}_{a} \times \hat{p}).\hat{s}
\end{equation}
For the central field approximation the Hamiltonian $\hat{H}^{SOC}$ [23] can be written as 
\begin{equation}
\hat{H}^{SOC} = \zeta\hat{l}.\hat{s}
\end{equation}
where $\hat{l}$ is the angular momentum and 
$\zeta = \frac{1}{2m^2c^2r}\frac{d\hat{V}_{a}}{dr}$, where $c$ is the speed of light.

The lattice information were taken from the materials research project site [24]. The lattice constants, the kinetic energy cut-off (Ecutwfc) and charge 
density cut-off (Ecutrho) values used for $\textrm{RB}_{4}$ are mentioned in Table 1. All the calculations were performed on three dimensional crystals consisting of primitive 
tetragonal lattice with 20 atoms. The energy conservation was achieved using $8^{3}$-points in the full Brillouin zone for sampling. Energy convergence criteria of $10^{-6}$ Ry 
were used for self-consistent calculations. The band structure is plotted along the path involving high symmetry points. The high symmetry points for tetragonal lattice system in 
the first Brillouin zone are  $\Gamma$=$(0,0,0)$, $X$= ($\frac{\pi}{a}$,0,0), $M$=($\frac{\pi}{a}$,$\frac{\pi}{a}$,0), $Z$=(0,0,$\frac{\pi}{c}$),  $R$=($\frac{\pi}{a}$,0,$\frac{\pi}{c}$),  
$A$=($\frac{\pi}{a}$,$\frac{\pi}{a}$,$\frac{\pi}{c}$). Calculated band structures were plotted along the high symmetry directions $\Gamma-X-M-\Gamma$, $Z-R-A-Z$, $X-R$, $M-A$.

\section{Results and Discussion}
In this study we have considered 4 canonical systems $\textrm{LaB}_{4}$, $\textrm{CeB}_{4}$, $\textrm{NdB}_{4}$ and $\textrm{SmB}_{4}$ with relatively small spin-orbit coupling strength and 4 
canonical systems $\textrm{HoB}_{4}$, $\textrm{ErB}_{4}$, $\textrm{TmB}_{4}$ and $\textrm{LuB}_{4}$ with much larger spin-orbit coupling effect. In Table 1. we have summarized the lattice constants 
for systems under consideration. In Table 2. we have summarized the atomic spin-orbit coupling energy [25](in units of $\textrm{cm}^{-1})$ of various rare-earth atoms under consideration.
\begin{table}[htbp]
\begin{center}
\begin{tabular}{|c|c|c|c|c|c|}
\hline 
Materials & a($\AA$) & c($\AA$) & Ecutwfc(Ry)& Ecutrho(Ry)\\
\hline
	$\textrm{LaB}_{4}$ & $7.31066$ & $4.18269$ & $25$ & $225$ \\
	$\textrm{CeB}_{4}$ & $7.17377$ & $4.07463$ & $40$ & $340$ \\
	$\textrm{NdB}_{4}$ & $7.23842$ & $4.11996$ & $38$ & $342$ \\
	$\textrm{SmB}_{4}$ & $7.18656$ & $4.08152$ & $35$ & $315$ \\ 
	$\textrm{HoB}_{4}$ & $7.08619$ & $4.00815$ & $42$ & $340$ \\ 
	$\textrm{ErB}_{4}$ & $7.06973$ & $3.99708$ & $37$ & $332$ \\
	$\textrm{TmB}_{4}$ & $7.05321$ & $3.98405$ & $38$ & $340$ \\
	$\textrm{LuB}_{4}$ & $7.02687$ & $3.96821$ & $42$ & $378$ \\
\hline
\end{tabular}
\end{center}
\caption{Lattice constants and the parameters used for the calculations.}
\end{table}
\begin{table}[htbp]
\begin{center}
\begin{tabular}{|c|c|c|c|c|}
\hline 
	Elements & SOC Energy($\textrm{cm}^{-1}$) & Elements & SOC Energy($\textrm{cm}^{-1}$)\\
\hline
	$\textrm{La}$ & $5.6\times 10^{3}$ & $\textrm{Ho}$ & $8.1\times 10^{3}$\\
	$\textrm{Ce}$ & $5.8\times 10^{3}$ & $\textrm{Er}$ & $8.4\times 10^{3}$\\
	$\textrm{Nd}$ & $6.3\times 10^{3}$ & $\textrm{Tm}$ & $8.7\times 10^{3}$\\
	$\textrm{Sm}$ & $6.8\times 10^{3}$ & $\textrm{Lu}$ & $9.3\times 10^{3}$\\

\hline
\end{tabular}
\end{center}
\caption{Tabulation for spin-orbit energy}
\end{table}
\begin{table}[htbp]
\begin{center}
\begin{tabular}{|c|c|c|}
\hline 
Materials &\multicolumn{2}{c|}{Fermi energy $E_{F}$(eV)}\\
\cline{2-3} 
 &Without SOC & With SOC \\
\hline
	$\textrm{LaB}_{4}$ & $12.447$ & $12.523$ \\
	$\textrm{CeB}_{4}$ & $13.103$ & $13.103$  \\
	$\textrm{NdB}_{4}$ & $12.290$ & $12.303$ \\
	$\textrm{SmB}_{4}$ & $12.285$ & $12.301$ \\

\hline

\end{tabular}
\end{center}
\caption{Fermi Energy due to effect of low spin-orbit interaction materials.}
\end{table}
It is important to mention that the choice of kinetic energy cut-off and the number of $k$-points chosen over the irreducible Brillouin zone are extremely crucial in determining crystal structure 
and band structure calculation. We have calculated the total energy as a function of the plane wave kinetic energy cut-off as well as the number of $k$-points over irrudicible Brillouin zone. 
In Fig. 2(a), (b) we show the convergence of the total energy as a function of kinetic energy cutoff and number of $k$-points for one canonical system $\textrm{LaB}_{4}$ with low spin-orbit coupling 
strength. In Fig. 2(c), (d) we have shown the same for $\textrm{LuB}_{4}$, a material with much larger SOC strength. It is clear that the kinetic energy cut-off in the range of 20-50 Ry and 30-50 Ry 
are deemed to be sufficient for convergence of total energy in these two systems, respectively. Also, we have found that $4\times 4\times 4$ $k$-mesh (defined over irreducible Brillouin zone) is 
sufficient for relative stability of tetragonal structure. For the entire calculation we have chosen a $k$-mesh of size $8\times 8\times 8$.
\begin{figure}
\centering
\includegraphics[scale=0.35]{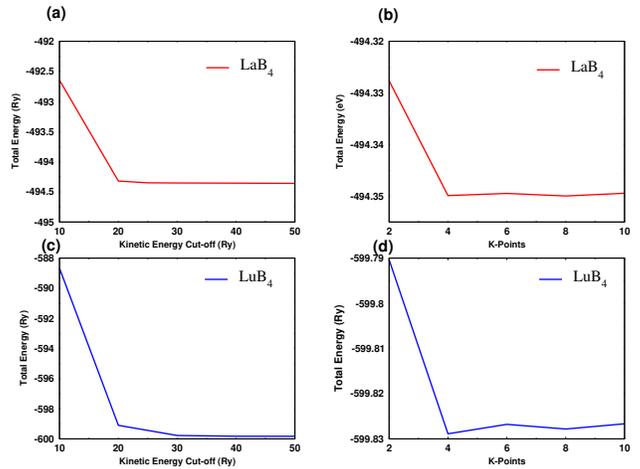} 
\caption{Total Energy as a function of kinetic energy cut-off and $k$-points along irreducible edges. Convergence of self-consistent field for (a,b) $\textrm{LaB}_{4}$ and (c,d) $\textrm{LuB}_{4}$, respectively.}
\end{figure}
\subsection{Systems with low SOC effect} 
Taking the optimized crystal structure, we have calculated the electronic band structures and projected density of states (PDOS) with and without spin-orbit coupling effects under generalized 
gradient approximations (GGA) and GGA+SO, respectively. In Table 3. we compare Fermi energy for systems with and with out SOC. The Fermi energy for $\textrm{LaB}_{4}$ changes significantly but for 
other systems change is only at the second decimal place. The main reason is that except for $\textrm{LaB}_{4}$ (with SOC) the pseudo-potentials in the non-magnetic state does not involve highly 
localised $4f$ orbitals and SOC strongly affects $4f$ orbitals and its effect on other orbitals are only secondary through hibridization with $4f$ orbitals. In Fig. 3(a) we have shown the band 
structure for $\textrm{LaB}_{4}$ with and without SOC effect. The Fermi level is set to zero for both the cases. As can be clearly observed from Fig. 3(a), (b) except at discrete symmetry points 
$\Gamma$, $Z$ and $R$ there is no significant SOC effect especially near the Fermi energy. However SOC lifts degeneracy at special symmetry points. Also, it can be observed that along the path $R-A$ 
bands are very flat and there is wide gap (of about 4 eV) between the top and bottom bands in this region. Flat bands correspond to non-dispersive localized bands arising mainly from deep core level 
state. In Fig. 4(a), (b) we have shown projected DOS from various orbitals at a given site in the absence of SOC. At the Fermi level the contribution is predominantly from B $2p$ and La $3d$. Discrete 
spectral peaks at -32 eV, -15 eV etc. arises due to deep core level states like  B $1s$, La $1s$, $2p$. In Fig. 4(c) we show combined PDOS from all atoms as well as total DOS. When we switch on SOC the 
B $2p$ state gets split into two peaks corresponding to $j=0.5$ and $j=1.5$. Also La $3p$ state gets split into two peaks to $j=0.5$ and $j=1.5$. In the presence of SOC there is contribution of $4f$ 
state (split into $j=2.5$ and $j=3.5$) at the Fermi energy. This is an unique feature in the case of $\textrm{LaB}_{4}$ and is absent in all other systems we have considered in this study. 
PDOS corresponding to $4f$ is spread over wide range of energy from -10 eV to 7 eV but the total spectral weight is much smaller than B $2p$ and La $3d$ contributions. Just above the Fermi level, in the 
range 0 to 7.5 eV, PDOS arises due to strong hybridization between La $3d$ orbitals and B $1s$, $2p$ orbitals.
\begin{figure}[htbp]
\centering
\includegraphics[width=0.90\linewidth, height=8.0cm]{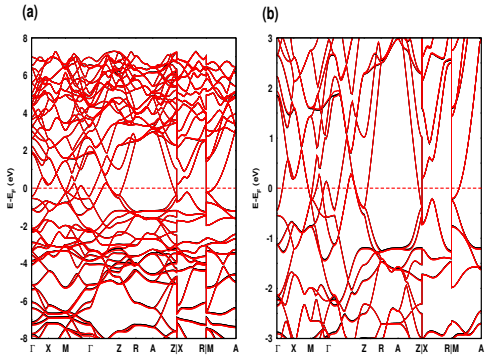} 
	\caption{(a) (Color online) Electronic band structure for $\textrm{LaB}_{4}$. (a) represents band structure without (black) and with SOC (red). (b) represents 
same band structure in the narrower energy window about the Fermi level (set to zero).}
\end{figure} 
 
\begin{figure}[htbp]
\centering
\includegraphics[width=0.90\linewidth, height=8.0cm]{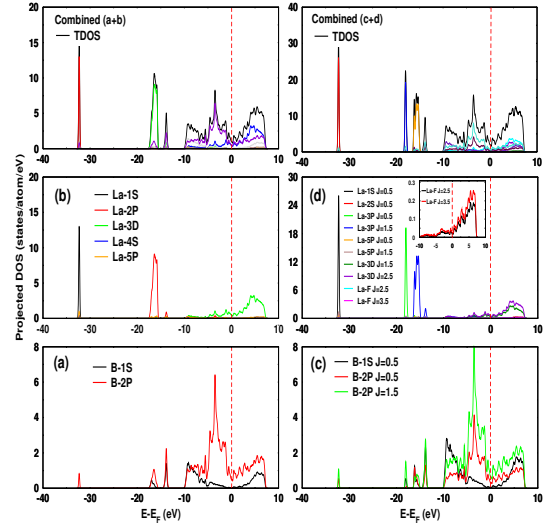} 
	\caption{ (c) PDOS of $\textrm{LaB}_{4}$ in the absence of SOC. (a) and (b) shows contribution from different orbitals from B and La, respectively. Top left panel shows combined contribution from all 
	orbitals. Dicrete peaks at -32 eV, -17 eV arises mainly due to La $1s$ and La $2p$ orbitals while the peak at -15 eV due to B $2p$. (c) and (d) shows spin-split contribution from varios B and La 
	orbitals in the presence of SOC. The peak at -17 eV gets split into two peaks with $j=0.5$ and $j=1.5$. Inset of (d) shows contribution of spin-split $4f$ orbitals about Fermi level. Top right panel 
	shows the combined contribution from all orbitals.}
\end{figure} 
 In Fig. 5 we summarize the band structure  and projected density of states of $\textrm{CeB}_{4}$ with and without SOC effects. $Ce$ is the first atom in the lanthanides series which contains $4f$ 
 orbital. As can be clearly seen from Fig. 5(a) and 5(b) in the presence of SOC, otherwise degenerate bands split at $\Gamma$ and R points but the bands remain degenrate at Z point. As in the case of 
 $\textrm{LaB}_{4}$ there exists non-dispersive flat bands along $R-A-Z$ directions and there is a gap of around 4.5 eV between the top and the bottom bands. In Fig. 5(c) we show the PDOS arising from 
 various atomic orbitals in the absence of SOC effects. The distinct spectral peaks appearing at -14 eV and at -17 eV are due to B $2p$ and Ce $3p$ orbitals, respectively. The extremely narrow spectral 
 peak at -34 eV arises due to deep core level Ce $1s$ state and B $2p$. The continuum density of states in the energy window -10 eV to 8 eV arises due to hybridized B $2p$ and Ce $3p$, Ce $4p$, Ce $5d$ 
 orbitals. In Fig. 5(d) we show the effect of SOC on PDOS for various atoms. As in the case of $\textrm{LaB}_{4}$ there is appearance of extremely narrow peak at -19 eV due to splitting of spin-degeneate 
 B $2p$ and Ce $3p$ orbitals into $j=0.5$ and $j=1.5$ manifolds. The DOS in the energy window -10 eV to 8 eV remains largely unaffected as in the case of $\textrm{LaB}_{4}$ and there is no additional 
 contribution due to spin split Ce $4f$ orbitals.  

\begin{figure}[htbp]
\centering
\includegraphics[width=0.90\linewidth, height=8.0cm]{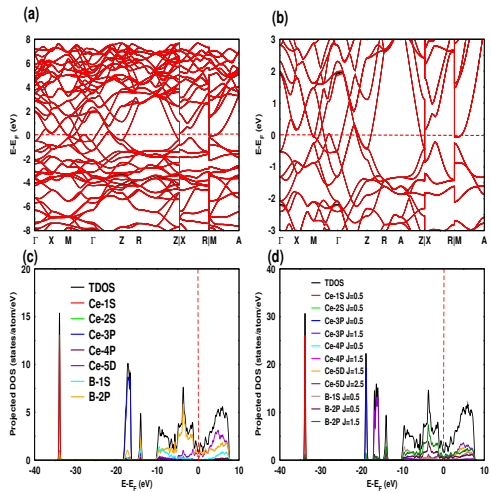} 
\caption{(a) (Color online) Combined Band structure of $\textrm{CeB}_{4}$ with out (black) and with SOC (red), (b) same band structure in the narrower energy window about the Fermi level (set to zero). Degenerate band splits at $\Gamma$ and R points. Fig. (c) and (d) The PDOS of $\textrm{CeB}_{4}$. The distinct spectral peaks appearing at -14 eV and at -17 eV are due to B $2p$ and Ce $3p$ orbitals, respectivelyIn the energy window -10 eV to 8 eV represents continuum DOS due to hybridized B $2p$ and Ce $3p$, Ce $4p$, Ce $5d$ orbitals.}
\end{figure}

 The band structure  and projected density of states of $\textrm{NdB}_{4}$ with and without SOC effects have been summarised in Fig. 6. As shown in Fig. 6(a) and 6(b), the spin degenerate bands 
 splits in various regions due to SOC effects present in these systems. Band splitting is more explicit along the direction $\Gamma - Z - R$ and $A-Z$. Very few bands cros the Fermi level and 
 far from Fermi level most of the bands are much less dispersed and nearly flat. In Fig. 6(c) we show the PDOS from various atoms without SOC effects. As in the earlier cases the distinct spectral 
 peak at -15 eV arises due to B $2p$ orbitals and the spectral peak at -19.5 eV arises due to B $2p$ and Nd $3p$ orbitals, respectively. The extremely narrow spectral peak at -38 eV arises mainly due to 
 non-dispersive deep core-level Nd $1s$ orbital. However B $2p$ orbitals have also contribution towards the peak at -38 eV. The continuum density of states in the energy range between -10.5 eV to 7.5 eV 
 arises due to bybridized B $2p$ and Nd $3p$, Nd $4p$, Nd $5d$ orbitals. Finally, in Fig. 6(d) we show the PDOS in the presence of SOC effect. The continuum DOS in the range -10.5 eV to 7.5 eV remains 
 largely unaffected. However the peak at -19.5 eV gets split into two peaks at -18 eV and -21 eV. This arises due to otherwise degenrate B $2p$ and Nd $3p$ orbitals splitting into $j=0.5$ and $j=1.5$ 
 manifolds due to SOC effects.

\begin{figure}[htbp]
\centering
\includegraphics[width=0.90\linewidth, height=8.0cm]{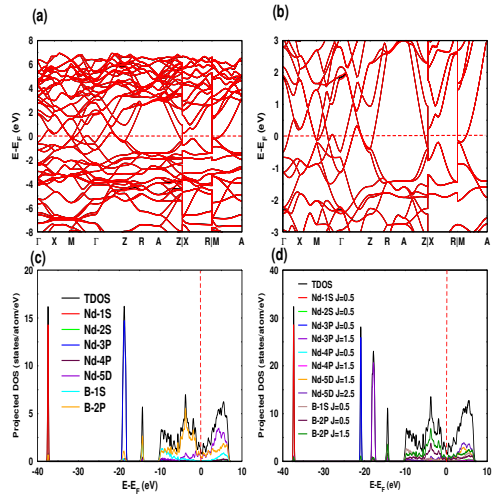} 
\caption{(a) (color online) Combined Band structure of  $\textrm{NdB}_{4}$ with out (black) and with (red) SOC, (b) same band structure in the narrower energy window about Fermi level (set to zero). Fig. (c) and (d) The PDOS of $\textrm{NdB}_{4}$. Very small narrow spectral peak at -38 eV is found due to non-dispersive deep core-level Nd $1s$ orbital. Other peaks at -38 eV arises for B $2p$ orbitals. The continuum DOS in the energy range between -10.5 eV to 7.5 eV arises due to bybridized B $2p$ and Nd $3p$, Nd $4p$, Nd $5d$ orbitals.}
\end{figure}
In Fig. 7 repregents the band structure and density of states of $\textrm{SmB}_{4}$ with and without SOC effects. It is interesting to mention that $\textrm{SmB}_{4}$ is metallic whereas  
$\textrm{SmB}_{6}$ is a Kondo insulator where Sm shows mixed valency $\textrm{Sm}^{+2}$ and $\textrm{Sm}^{+3}$ at the ratio 3:7. In Fig. 7(a) and 7(b) we show electronic band structure. 
Splitting of energy bands in the $\Gamma-Z-R$ direction is much more prominent due to much larger SOC effects. Energy bands along $R-A-Z$ continues to remain flat. In Fig. 7(c) we show PDOS. The 
discrete peak arising due to Sm $1s$ shifts further down to -41 eV. The spectral peak at -20.5 eV and -14.5 eV arises due to B $2p$, Sm $3p$ and B $1s$, $2p$ orbitals, respectively. The origin of 
continuum states in the range -10.5 eV to 8 eV is same as earlier. When we switch on the SOC the spectral peak arising due to $p$-orbitals of B and Sm gets split into $j=0.5$ and $j=1.5$ states and 
the corresponding spectral peaks apears at -23 eV and -19 eV, respectively.     
\begin{figure}[htbp]
\centering
\includegraphics[width=0.90\linewidth, height=8.0cm]{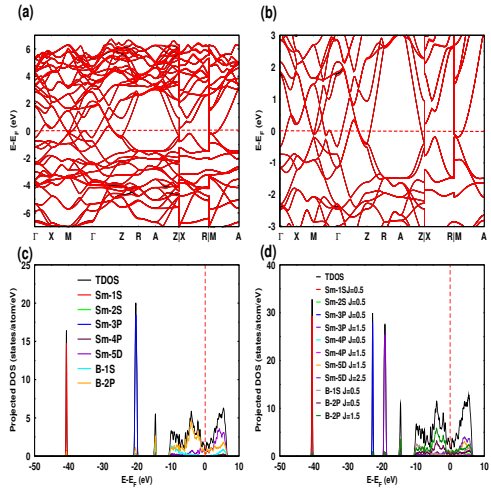}
\caption{(a) (color online) Combined Band structure of $\textrm{SmB}_{4}$ with out (black) and with (red) SOC, (b) same band structure in the narrower energy window about the Fermi level (set to zero). Fig. (c) and (d) the PDOS of $\textrm{SmB}_{4}$. The isolated spectral peak at -20.5 eV and -14.5 eV corresponds to B $2p$, Sm $3p$ and B $1s$, $2p$ orbitals, respectively.} 
\end{figure}
\subsection{Systems with large SOC Effect}
In the previous section we  have considered SOC effects on 4 canonical systems with relatively low SOC effect. In this section we consider SOC effects on 4 canonical systems with relatively large SOC effects. 
In Table. IV we have summarized the Fermi energy with and without SOC effects.

In Fig. 8 we have shown the band structure and projected DOS for $\textrm{HoB}_{4}$ with and without SOC effects. SOC effects on the splitting of energy bands are prominent for wide range of energies. Bands far away 
from Fermi energy are also affected due to strong SOC effects. Degeneracy lifting effect along $\Gamma-Z-R$ are now quite explicit. Fermi level crossing bands along $Z-R$ are also affected. However flat bands along 
$R-A-Z$ are not affected by SOC. In Fig. 8(c) we show PDOS due to various atoms as earlier. Continuum DOS in the range -10.5 eV to 7.5 eV arizes due to strong hybridization between $5d$ orbitals of Ho and $2p$ orbitals 
of B atoms. The spectral peak at -14 eV arises due to $2p$ orbitals of B atoms while the peak at -24 eV arises due to Ho $3p$ orbitals. Extremely narrow and isolated peak at -48 eV arises due to deep core level $1s$ 
orbital of Ho atom. In Fig. 8(d) we show the effect of SOC on PDOS. There is enhancement of PDOS around Fermi level. The spectral peak at -24 eV gets split into two peaks at -27 eV and -22 eV which arises due to Ho 
orbitals with $j=0.5$ and $j=1.5$, respectively. 

	Fig. 9 indicates the band structure  and projected density of states of $\textrm{ErB}_{4}$ in the presence and absence of SOC effects. As seen in Fig. 9(a) and 9(b), spin-split bands are quite visible in the 
energy range -4 eV to -6 eV along $\Gamma-Z-R$ direction. Band splitting effects near the Fermi level also starts showing up. Projected density of states as shown in Fig. 9(c) follows similar trend as in the 
case of other tetra-borides. The continuum density of states in the range -10 eV to 7 eV arises from the hybridized B $2p$ and Er $5d$ orbitals. The spectral peak due to Er $1s$ is now at -51 eV. While the Er $3p$+ B $2p$ 
spectral peak is at -25 eV. The smaller peak arising due to B $2p$ is at -15 eV. As shown in Fig. 9(d), inclusion of SOC effect causes splitting of the -25 eV peak into $j=0.5$ and $j=1.5$ states situated at 
-29 eV and -23.5 eV, respectively.
\begin{table}
\begin{center}
\begin{tabular}{|c|c|c|}
\hline 
Materials &\multicolumn{2}{c|}{Fermi energy $E_{F}$(eV)} \\
\cline{2-3} 
 &Without SOC & With SOC \\
\hline
	 $\textrm{HoB}_{4}$ & $12.271$ & $12.279$\\
	 $\textrm{ErB}_{4}$ & $12.280$ & $12.286$\\
	 $\textrm{TmB}_{4}$ & $12.274$ & $12.278$ \\
	 $\textrm{LuB}_{4}$ & $12.251$ & $12.251$ \\
\hline
\end{tabular}
\end{center}
\caption{Fermi Energy due to effect of large spin-orbit interaction materials.}
\end{table}
\begin{figure}[htbp]
\centering
\includegraphics[width=0.90\linewidth, height=8.0cm]{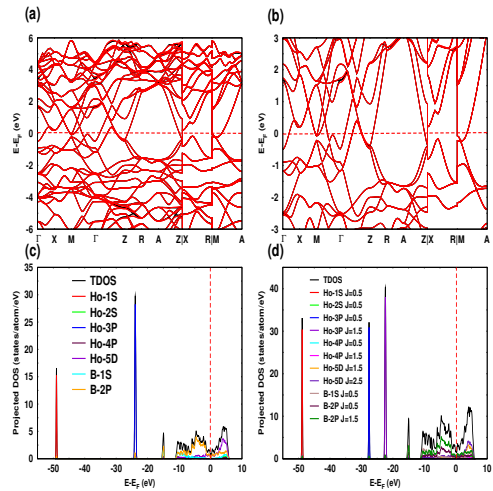} 
\caption{(a) (Color online) Combined Band structure of $\textrm{HoB}_{4}$ without (black) and with (red) SOC, (b) same band structure in the narrower energy window about the Fermi level. The spectral peak at -14 eV corresponds to $2p$ orbitals of B atoms while the peak at -24 eV arises due to Ho $3p$ orbitals. Extremely narrow and isolated peak is found at -48 eV due to deep core level $1s$ 
orbital of Ho atom. Fig. (c) and (d) The PDOS of $\textrm{HoB}_{4}$. Showing the individual contributions from each orbital atom with and without SOC effect. The Fermi energy is set to zero assigned by the dotted red line.}
\end{figure}

\begin{figure}[htbp]
\centering 
\includegraphics[width=0.90\linewidth, height=8.0cm]{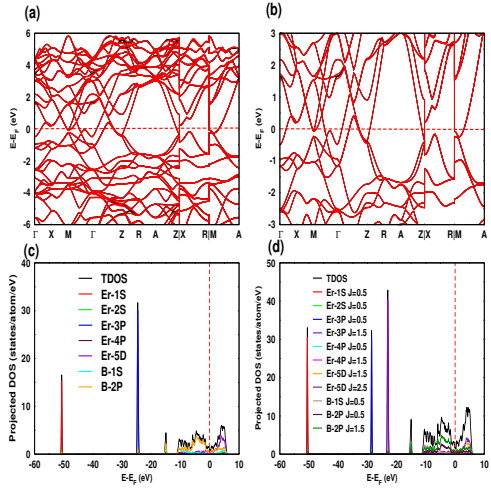} 
\caption{(a) (Color online) Combined Band structure of $\textrm{ErB}_{4}$ with (red) and without SOC (black), (b) same band structure in the narrower energy window about the Fermi level (set to zero). The spin-split bands appearing in the energy range -4 eV to -6 eV along $\Gamma-Z-R$ direction. (c) and (d) PDOS of $\textrm{ErB}_{4}$. The spectral peak due to Er $1s$ is now at -51 eV. While the Er $3p$+ B $2p$ spectral peak is at -25 eV . At -15 eV due to B $2p$ orbital. (d) Inclusion of SOC effect causes splitting of the -25 eV peak into $j=0.5$ and $j=1.5$ states situated at -29 eV and -23.5 eV, respectively}.
\end{figure}
In Fig. 10 we summarize the electronic band structure and projected density of states of $\textrm{TmB}_{4}$. In an earlier study band structure for $\textrm{TmB}_{4}$ in the anti-ferromagnetic state was reported. So the 
present study is relevant in the paramagnetic state of this system. As shown in Fig. 10(a) and 10(b) energy bands far from the Fermi level are strongly affected due to SOC. Energy bands in the energy range -4 eV to -6 eV 
show significant splitting especially along $\Gamma-Z-R$ direction. Similar features are also observable for energy bands in the window 1 eV to 2 eV. Some of the Fermi level crossing bands show degeneracy lifting effects near 
Fermi level. The spectral features are similar to the other tetraborides. The peaks arising due to Tm $1s$, $3p$ and B $2p$ are at -52.5 eV and -26 eV respectively. The continuum DOS in the energy range -11 eV to 7 eV arises 
due to hybridized Tm $4p$, $5d$ orbitals with B $2p$ orbitals. Inclusion of SOC, as shown in Fig. 10(d), causes splitting of the -26 eV spectral peaks into a $j=0.5$ peak at -30 eV and a $j=1.5$ peak at -24 eV.   

Finally, in Fig. 11 we show our results for $\textrm{LuB}_{4}$. Incidentally Lu is the last member of the lanthanide series with completely filled $4f$ orbitals. As in the case of $\textrm{TmB}_{4}$ there is strong SOC 
effects on the energy bands in the energy window -6 eV to -4 eV as well as in the window 1 eV to 2 eV. SOC effects on the Fermi level crossing bands near Fermi energy are less compared to $\textrm{TmB}_{4}$. These features 
are well summarised in Fig. 11(a) and 11(b). In Fig. 11(c) and 11(d) we show the projected DOS in the absence and presence of SOC effects, respectively. The spectral peak at -56 eV is due to Lu $1s$ orbital while the peak 
at -15 eV is due to B $1s$ and $2p$. There is a strong peak at -27 eV arising due to Lu $3p$ orbital. The height of this peak is much more than the other two discrete peaks. The continuum of density of states around Fermi 
level arises due to hybridized B $2p$ and Lu $5d$ orbitals. In the presence of SOC the peak at -26 eV gets split onto two peaks at -32 eV and -25 eV with $j=0.5$ and $j=1.5$, respectively.

\begin{figure}[htbp]
\centering
\includegraphics[width=0.90\linewidth, height=8.0cm]{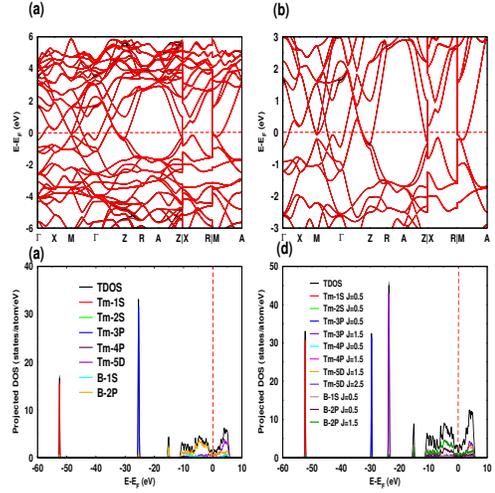} 
\caption{(a) (color online) Combined Band structure of $\textrm{TmB}_{4}$ without (black) and with (red) SOC, (b) same band structure in the narrower energy window about Fermi level (set to zero). Fig. (c) and (d) The PDOS of  $\textrm{TmB}_{4}$. Showing the individual contributions from each orbital atom with and without SOC effect. The continuum DOS in the energy range -11 eV to 7 eV arises 
due to hybridized Tm $4p$, $5d$ orbitals with B $2p$ orbitals. With SOC, the p-orbital split into a $j=0.5$ peak at -30 eV and a $j=1.5$ peak at -24 eV. }
\end{figure}
\begin{figure}
\centering 
\includegraphics[width=0.90\linewidth, height=8.0cm]{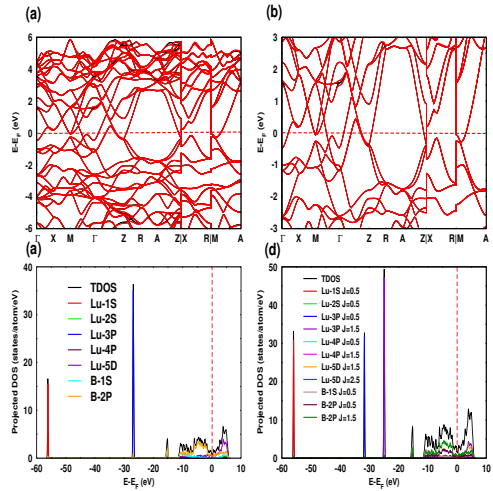} 
\caption{(a) (color online) Combined Band structure of  $\textrm{LuB}_{4}$ without (black) and with (red) SOC, (b) same band structure in the narrower energy window about Fermi level (indicates dotted red line). Fig. (c) and (d) The PDOS of $\textrm{LuB}_{4}$. In the presence of SOC the peak at -26 eV gets split onto two peaks at -32 eV and -25 eV with $j=0.5$ and $j=1.5$, respectively like other system.}
\end{figure}
\section{Conclusion}
	We have investigated the electronic structure of $\textrm{RB}_{4}$ with non-magnetic ground state. The electronic
band structure shows splitting due to interaction between spin and angular momentum. The bands splitting has been interpreted with the help of PDOS. It has also been observed that the two new branches for p-orbital appearing due to SOC effect. In case of $\textrm{LaB}_{4}$ with SOC, the contribution of $4f$ orbitals to the DOS about the Fermi level has been observed. This work is partially supported by WB-
DSTBT research grant no. STBT- 11012(26)/31/2019-ST SEC. One of us (NP) would like to acknowledge
hospitality of IIT, Kharagpore. One of us (IS) would like to thank Bajkul Milani Mahavidyalaya (College) authority for giving me an
opportunity to pursue research as a Ph. D. scholar.

\end{document}